\documentstyle[fleqn,11pt]{article}
\begin{document}
\baselineskip0.26in
\title{
\hskip5.2in\parbox[t]{1.25in}{\large\raggedright NS--USTC/96-16 
}\\
\vskip1.25cm
Higher Order Perturbation Corrections of Rotating Excited States
in the Standard SU(3) Skyrme Model to Baryon Mass Spectrum 
\thanks{This work was supported by National Natural Science Foundation
of China through C.N.Yang and Science Fund of the Chinese Academy of
Science.}}
\author{
{\large Jin-Ping Li, Mu-Lin Yan and Rong-Hua Yu}\\
Fundamental Physics Center and Nonlinear Science Center,\\ 
University of Science \& Technology of China,\\
Hefei, Anhui, 230026, P.R.China}
\date{June 1}

\maketitle
\vskip3.0cm 
\begin{abstract}
{ The higher order corrections of SU(3) rotating excited states
to the Gell-Mann-Okubo Relations (or GOR) are presented in the 
standard SU(3) Skyrme model.
The Improved GOR (or IGOR) are obtained. 
The results show the IGOR for decuplet up to the third order and 
for octet up to the second order are much compatible with 
the experimental data. But things becomes quite inadequate
for the octet to the third order. In order to overcome 
the inadequacy, a heuristic discussion is presented. 
The properties of SU(3) rotating excited states 
27-let(with spin $\frac{1}{2}$ or $\frac{3}{2}$),10$^*$-let
(with spin $\frac{1}{2}$) and 35-let(with spin $\frac{3}{2}$) 
are also discussed.}
\end{abstract}
\vskip0.05cm
\newpage
\section{Introduction}

 With the extensive and intensive study of long-range 
QCD, Skyrme model$^{[1-10]}$ or the chiral soliton model is believed
to be a promising theory to depict the behavior of long-range strong 
interaction. In this theory, the baryon-multiplets emerge as topological 
solitons (i.e.,skyrmions) in the SU(3)$\otimes$SU(3) current algebra 
chiral Lagrangians. These solitons can be excited through the
quantization of rotating fluctuations in terms of collective coordinates.
It is believed that this chiral soliton model provides a reasonable 
dynamical mechanism for the mass-splitting of SU(3)-baryons.
Therefore it has been applied extensively to the study of the mass 
relations for baryons$^{[3-10]}$. 

Among these relations, the most famous ones are Gell-Mann-Okubo 
Relations (GOR in short)$^{[11]}$, which were originally 
formulated in terms of a perturbative treatment of flavor-breaking 
in SU(3) group theory. However, there are deviations between 
GOR and experimental data, which can be shown as follows,
\begin{equation}
\label{GOR10}
M_{\Sigma^{*}}-M_{\Delta}=
M_{\Xi^{*}}-M_{\Sigma^{*}}+\delta m_{10}^{(1)}=
M_{\Omega}-M_{\Xi^{*}}+\delta m_{10}^{(2)}.\\
\end{equation}
\begin{equation}
\label{GOR8}
2(M_{N}+M_{\Xi})=3M_{\Lambda}+M_{\Sigma}+\delta m_{8},\\
\end{equation}
where $\delta m_{10}^{(1)},\delta m_{10}^{(2)},\delta m_{8} $
stand for the deviations.
As $\delta m_{10}^{(1)}=\delta m_{10}^{(2)}=\delta m_{8}=0,$
eqs.(1) and (2) go back to the standard GOR. However, in the real
world, $\delta m_{10}^{(1)},\delta m_{10}^{(2)},\delta m_{8}$ are not
zero$^{[12]}$,
\begin{eqnarray}
\label{GOR101}
\delta m_{10}^{(1)} &=& 3.8MeV,\\
\label{GOR102}
\delta m_{10}^{(2)} &=& 13.6MeV,\\
\label{GOR81}
\delta m_{8} &=& -26.1MeV,
\end{eqnarray}

In order to re-establish GOR and especially reveal the deviations,
the deliberate consideration of the flavor breaking term is needed. 
Generally, three methods are often used to deal with 
the mass-splitting in the SU(3) Skyrme model, which are 
perturbative$^{[3-8]}$, non-perturbative$^{[9]}$ and 
bound-state approach$^{[10]}$, respectively. In ref.[5] the standard 
GOR has been actually re-established through one order 
perturbative calculations in the SU(3) Skyrme model, 
which could be regarded as a refurbished version of the original
perturbative treatment in SU(3) group theory. 
Therefore it is reasonable to expect that higher order perturbative
corrections in SU(3) skyrmion quantum mechanics could further improve
the standard GOR. 
In ref.[6], the perturbative calculations have been carried out 
for octet-baryons up to the third order.
The similar calculations for decuplet, however, are left due to the 
lack of the corresponding Clebsh-Gordon Coefficients (CGC's in short) 
for SU(3) 
group in the literatures$^{[13,14]}$. A systematical study of GOR 
and its improvement through the skyrmion quantum mechanics remain also 
to be open. In the present paper, 
the Gel'fand bases are used to calculate the CGC's for SU(3) directly,
and the Improved GOR (IGOR in short) are derived for both decuplet and 
octet up to the second order, then the deviations of 
of $\delta m_{10}^{(1)},\delta m_{10}^{(2)}$ and $\delta m_{8}$ are 
revealed which are quite compatible with experimental data. Up to the
third order, $\delta m_{10}^{(1)},\delta m_{10}^{(2)}$ are found to be
more positive, but unfortunately $\delta m_{8}$ to be much inadequate.
Therefore we expect that further study on this problem could 
remove the inadequacy.

This paper is organized as follows. 
In Sec.2, the formalism of the calculations is presented. 
In Sec.3, the calculations of SU(3) Clebsh-Gordon Coefficients 
are given. To the best of our knowledge, we find 
that there are not the CGC's of 8$\otimes$35 $\rightarrow$ 35 in 
the literatures$^{[13-14]}$ so far. These CGC's are listed in
the Appendix. The method employed to compute SU(3) CGC's in this 
paper is practical and convenient, and is also useful for other purpose.
In Sec.4, the calculations for higher order perturbation corrections 
to GOR of decuplet- and octet-baryons are presented, then the
IGOR's are obtained. In Sec.5, the comments and conclusions are given. 
A heuristic prescription is proposed which is expected to remove the 
inadequacy, then the role of the rotating excited states 
(or exotic SU(3) states) is discussed. 
\section{Formalism of Skyrmion Quantum Mechanics}

In the present paper, the notation of ref.[7] will be 
adopted. The Hamiltonian of the standard SU(3) skyrmion quantum 
mechanics in the collective coordinate space is$^{[7]}$ 
\begin{eqnarray}
\label{sumH}
H &=& H_{0}+H^{\prime}\\
\label{H0}
H_{0} &=& M_{s}+ \frac{1}{2b^{2}} ( \sum_{i=1}^{8}
L_{i}L_{i}-R_{8}^{2})+\frac{1}{2}(\frac{1}{a^{2}}-\frac{1}{b^{2}})
\sum_{A=1}^{3} R_{A}R_{A}+\frac{2 \delta}{\sqrt{3}} F_{\pi}R_{8}  \\
\label{H'}
H^{\prime} &=& m (1-D_{8 8}^{(ad)}(A))
\end{eqnarray}
where $M_{s}$ is the mass of classical soliton, $a^{2}$ and $b^{2}$ are the
soliton's ``moments of inertia'', $ m, F_{\pi} $ and
$\delta$ are constants and parameters in the model, $D_{\mu
\nu}^{(ad)}$ denotes the regular adjoint representation functions of
SU(3), and $[L_{i},L_{j}]=if_{ijk}L_{k},[R_{i},R_{j}]=-if_{ijk}R_{k},
[L_{i},R_{j}]=0.$ In eq.(~\ref{sumH}), $H_{0}$ serves as the unperturbed 
Hamiltonian, and $H^{\prime}$ as the perturbative part. It is easy to
see that $H_{0}$ is diagonal and the eigen-wavefunctions for $H_{0}$
are $^{[7]}$
\begin{equation}
\label{wavefunction}
|_{\mu \nu}^{\lambda} \rangle = (-1)^{s+s_{z}}\sqrt{\lambda}
D_{\mu,\nu}^{(\lambda)}(A)
\end{equation}
where $\lambda$ represents the Irreducible Representation (IR in short)
of SU(3), and $\mu =\left (\begin{array}{cc} I Y \\ I_{z}\end{array} \right ),
\nu =\left (\begin{array}{cc} S 1 \\ -S_{z}\end{array} \right )$ with
$I,Y,S$ denote isospin,hypercharge and spin respectively. 
The mass of the baryon for $| k\rangle \equiv |_{\mu \nu}^{\lambda}
\rangle $ can be calculated in perturbation,
\begin{equation}
\label{Eseries}
M_{k}=E_{k}^{(0)}+E_{k}^{(1)}+E_{k}^{(2)}+E_{k}^{(3)}+\cdots
\end{equation}
where
\begin{eqnarray}
\label{Ek0}
E_{k}^{(0)} &=& \langle k| H_{0} |k \rangle, \\
\label{Ek1}
E_{k}^{(1)} &=& \langle k| H^{\prime} |k \rangle, \\
\label{Ek2}
E_{k}^{(2)} &=& \sum_{n\not= k} \frac{|\langle n| H^{\prime} |k \rangle
|^{2}}{ E_{k}^{(0)}-E_{n}^{(0)}}\\
\label{Ek3}
E_{k}^{(3)} &=& \sum_{m\not= k} (\sum_{n\not= k} \frac{
\langle k| H^{\prime}|m \rangle \langle m| H^{\prime}|n \rangle 
\langle n| H^{\prime}|k \rangle}{ (E_{k}^{(0)}-E_{n}^{(0)})
(E_{k}^{(0)}-E_{m}^{(0)}) } - \frac{
\langle k| H^{\prime}|m \rangle \langle m| H^{\prime}|k \rangle
\langle k| H^{\prime}|k \rangle}{(E_{k}^{(0)}-E_{m}^{(0)})^{2}})
\end{eqnarray}
considering that spins of $|k \rangle$ and
 $|n \rangle$ are equal, the terms in the denominator of eqs.(~\ref{Ek2}),
(~\ref{Ek3}) can
 be obtained by considering eqs.(~\ref{H0}) and (~\ref{H'}),i.e.,
\begin{equation}
\label{Ek0-En0}
E_{k}^{(0)}-E_{n}^{(0)}=\frac{1}{2b^{2}}(C_{2}(k)-C_{2}(n)),
\end{equation}
where $C_{2}(k)$ denotes the Casimir operator for the $k$-dimensional
IR's of SU(3).
$E_{k}^{(0)}$ and $E_{k}^{(1)}$ have been known in ref.[7]. 
Our object is to calculate $E_{k}^{(2)}$ and $E_{k}^{(3)}$ which are
related to the non-diagonal matrix elements of $D_{8,8}^{(8)}(A)$
between the baryon-states (to see eqs.(~\ref{Ek2}),
(~\ref{Ek3})). 
From eqs.(~\ref{Ek1})-(~\ref{Ek3}), it can be easily seen that the crux of
the calculations is to deal with such matrix element as
$\langle m| H^{\prime}|n \rangle$. According to ref.[7,9],
and noting that $D_{8,8}(A)=D_{
\left (\begin{array}{cc} 0 0 \\ 0\end{array} \right )
\left (\begin{array}{cc} 0 0 \\ 0\end{array} \right )}^{8},$
the matrix element can be calculated with the following formula,
\begin{equation}
\label{matrixelement}
\langle_{\mu_{2} \nu_{2}}^{\lambda_{2}}| D_{88}^{(8)}(A)|_{\mu_{1} 
\nu_{1}}^{\lambda_{1}}
\rangle=(-1)^{s_{1}+s_{2}+s_{1z}+s_{2z}}
\sqrt{\frac{\lambda_{1}}{\lambda_{2}}} \sum_{\gamma} 
\left ( \begin{array}{ccc}
\lambda_{1} & 8 & \lambda_{2\gamma} \\  \mu_{1} & \mu & \mu_{2}
\end{array} \right ) 
\left ( \begin{array}{ccc}
\lambda_{1} & 8 & \lambda_{2\gamma} \\  \nu_{1} & \nu & \nu_{2}
\end{array} \right )
\end{equation}
with a standard notation for the CGC's. Here $\mu=
\left (\begin{array}{cc} 0 0 \\ 0\end{array} \right )$ , $\nu=
\left (\begin{array}{cc} 0 0 \\ 0\end{array} \right )$ ,
$\gamma$ distingushes the independent IR's
occurring in the reduction $(8) \otimes (\lambda_{1})\rightarrow
(\lambda_{2})$. 
\vskip0.3in
\section{The calculations of SU(3) Clebsh-Gordon Coefficients}

In order to calculate the SU(3) CGC's, the choice of basis for
representations is of great importance. In the present section,
we will employ Gel'fand basis$^{[15,16]}$ and confine ourselves
to discussing the calculations of only SU(3) CGC's.

In terms of the Gel'fand symbol, any basis vector of an IR of SU(3)
can be labelled as
\begin{eqnarray}
\label{basis}
\left (\begin{array}{c} \lambda \\ \mu\end{array} \right) \equiv
\left (\begin{array}{c}
       \begin{array}{ccc} m_{13}& m_{23}& m_{33} \end{array}\\
       \begin{array}{cc} m_{12} & m_{22} \end{array}\\
       \begin{array}{c} m_{11} \end{array}
       \end{array} \right)
\end{eqnarray}
where the parameters $m_{i,k}$'s stand for the partitions of the IR's
of SU(N) ($N \leq 3$), e.g.,($m_{13},m_{23},m_{33}$) correspond to
the IR's of SU(3).

The matrix element of SU(3) generator $E_{32}$ between the above
basis is$^{[15,16]}$
\begin{equation}
E_{32}\left (\begin{array}{c}
       \begin{array}{ccc} m_{13} & m_{23}& m_{33} \end{array}\\
       \begin{array}{cc} m_{12} & m_{22} \end{array}\\
       \begin{array}{c} m_{11} \end{array}
       \end{array} \right)= A
       \left (\begin{array}{c}
       \begin{array}{ccc} m_{13}& m_{23}& m_{33} \end{array}\\
       \begin{array}{cc} m_{12}-1 & m_{22} \end{array}\\
       \begin{array}{c} m_{11} \end{array}
       \end{array} \right)+ B
       \left (\begin{array}{c}
       \begin{array}{ccc} m_{13}& m_{23}& m_{33} \end{array}\\
       \begin{array}{cc} m_{12} & m_{22}-1 \end{array}\\
       \begin{array}{c} m_{11} \end{array}
       \end{array} \right)
\end{equation}

where
\begin{eqnarray}
A=(\frac{m_{11}-m_{12}}{m_{22}-m_{12}-1})^{\frac{1}{2}}
       ((-1) \frac{(m_{13}-m_{12}+1)(m_{23}-m_{12})(m_{33}-m_{12}-1)}
       {m_{22}-m_{12}})^{\frac{1}{2}}\\ \nonumber \\
B=(\frac{m_{11}-m_{22}+1}{m_{12}-m_{22}+1})^{\frac{1}{2}}
       ((-1) \frac{(m_{13}-m_{22}+2)(m_{23}-m_{22}+1)(m_{33}-m_{22})}
       {m_{12}-m_{22}+2})^{\frac{1}{2}}
\end{eqnarray}

 In the fundamental representation, the commutation relations of SU(3)
 generators $E_{ij}$ are
\begin{equation}
[E_{ij},E_{kl}]= \delta_{jk} E_{il}- \delta_{il} E_{kj}
\end{equation}
where $(E_{ij})_{\alpha \beta}=\delta_{i\alpha} \delta_{j\beta}-
\frac{1}{3} \delta_{ij} \delta_{\alpha\beta}$. Obviously, the matrix 
element of $E_{32}$ is always non-negative, because the generalized 
Condon-Shortley phase convention is followed, i.e., the matrix elements
of $E_{n,n-1}$ of SU(N) are defined to be non-negative.

 The SU(3) CGC's can be constructed with SU(3) isoscalar factors and
 SU(2) CGC's $^{[13]}$, i.e.,
\begin{equation}
\label{Wigner}
C_{CG}[SU(3)]=
\left(\begin{array}{cccc}
 \lambda_{1} & \lambda_{2} & | & \lambda \\
 I_{1}Y_{1} & I_{2} Y_{2} & | & I Y 
 \end{array} \right) C_{CG}[SU(2)]
\end{equation}
where $I,Y,\lambda$ represent isoscalar,hypercharge and dimension of
the IR of SU(3), respectively.

Then the calculations of SU(3) CGC's can be ascribed to the calculations
of SU(3) isocalar factors 
$\left(\begin{array}{cccc}
 \lambda_{1} & \lambda_{2} & | & \lambda \\
 I_{1}Y_{1} & I_{2} Y_{2} & | & I Y 
 \end{array} \right)$.
 Making use of the above formulas, the isoscalar factors of
 8 $\otimes$ 35 $\rightarrow$ 35 can be easily obtained, which are
 listed in the Appendix. All of the isoscalar factors have been 
 checked carefully so that the unitary and orthogonal conditions
 are preserved. The phase conventions are also checked on which
 the comments are given in Appendix.
\section{Higher order corrections to decuplet and octet}

\subsection{The baryon-decuplet sector}
   
 Considering the following decomposed formula of direct 
 product of the SU(3) IR's  
\begin{equation}
8 \otimes 10 = 35 \oplus 27 \oplus 10 \oplus 8
\end{equation}
we could see that the rotating excited states that acted as the 
intermediate states $|n \rangle$ and $|m \rangle$ in eqs.(~\ref{Ek2}) 
and (~\ref{Ek3}) can only take the following set
\begin{equation}
\{|m \rangle,|n \rangle\} \in \{ 35, 27, 8 \}.
\end{equation}
Making use of eq.(~\ref{matrixelement}) and eq.(~\ref{Wigner}), we 
could obtain all the non-zero
matrix elements for decuplet (see Table {\bf 1}.) 

  On the other hand, noticing $C_{2}(8)=3, C_{2}(10)=6,
C_{2}(27)=8$ and $C_{2}(35)=12$, we have
\begin{eqnarray}
E_{10}^{(0)}-E_{8}^{(0)} &=&\frac{3}{2b^{2}} \\
E_{10}^{(0)}-E_{27}^{(0)} &=&-\frac{1}{b^{2}}\\
E_{10}^{(0)}-E_{35}^{(0)} &=&-\frac{3}{b^{2}}
\end{eqnarray}

Then the complete results of the higher order corrections to the 
baryons in decuplet can be obtained from eqs.(~\ref{Ek0})-(~\ref{Ek3}) 
as follows
\begin{eqnarray}
\label{decupeq1}
M_{\Delta} &=& M_{10} - \frac{1}{8} m -\frac{85}{672} m^{2}b^{2} 
	-\frac{340}{129024} m^{3} b^{4}\\
\label{decupeq2}
M_{\Sigma^{*}} &=& M_{10} - \frac{26}{336} m^{2}b^{2} 
	-\frac{26}{8064} m^{3} b^{4}\\
\label{decupeq3}
M_{\Xi^{*}} &=& M_{10} + \frac{1}{8} m -\frac{9}{224} m^{2}b^{2} 
	-\frac{35}{14336} m^{3} b^{4}\\
\label{decupeq4}
M_{\Omega} &=& M_{10} + \frac{1}{4} m -\frac{5}{336} m^{2}b^{2} 
	-\frac{5}{4032} m^{3} b^{4}
\end{eqnarray}
where $M_{10}=\langle H_{0} \rangle _{\lambda =10}+m$. It is easy
to see that there are four non-linear equations and three unknown parameters.
Therefore the mass relation of the baryons in decuplet can be obtained
by solving the over-determinant equations.

\newpage
\centerline{{\bf Table 1. Values of non-zero matrix 
elements of $H'_{m,n}$ 
for decuplet}}

\begin{center}
\begin{tabular}{|c|cccccc|} \hline 
baryons & $H'_{10,10}/m$ & $H'_{27,27}/m$ & $H'_{35,35}/m$ 
& $H'_{10,35}/m$ & $H'_{10,27}/m$ & $H'_{27,35}/m$ \\ \hline
&&&&&&\\
$\Delta$ & $\frac{7}{8}$ & $\frac{99}{112}$ & $\frac{13}{16}$ & 
$\frac{5}{16}\sqrt{\frac{2}{7}}$ & $\frac{\sqrt{30}}{16}$ & 
$\frac{5}{16}\sqrt{\frac{3}{35}}$\\
&&&&&&\\
$\Sigma^{*}$ & 1 & $\frac{57}{56}$ & $\frac{7}{8}$ & 
$\frac{1}{4}\sqrt{\frac{5}{7}}$ & $\frac{1}{4}$ & 
$\frac{5}{8}\sqrt{\frac{1}{35}}$\\
&&&&&&\\
$\Xi^{*}$ & $\frac{9}{8}$ & $\frac{129}{112}$ & $\frac{15}{16}$ & 
$\frac{3}{16}\sqrt{\frac{10}{7}}$ & $\frac{\sqrt{6}}{16}$ & 
$\frac{5}{16}\sqrt{\frac{3}{35}}$\\
&&&&&&\\
$\Omega$ & $\frac{5}{4}$ & 0 & 1 & $\frac{1}{4}\sqrt{\frac{5}{7}}$ & 0 & 0\\ 
&&&&&&\\  \hline
\end{tabular}
\end{center}

 It is straightforward that the standard GOR for decuplet can be recovered
at the first order, i.e.,
\begin{equation}
\label{decuporder1}
M_{\Sigma^{*}}-M_{\Delta}=M_{\Xi^{*}}-M_{\Sigma^{*}}=M_{\Omega}-M_{\Xi^{*}}
\end{equation}
At the second order, the following constrained mass relation
can be obtained as the above over-determinant equations are solved,
\begin{equation}
\label{decuporder2}
M_{\Omega}-M_{\Delta}=3(M_{\Xi^{*}}-M_{\Sigma^{*}})
\end{equation}
eq.(~\ref{decuporder2}) is the IGOR for decuplet, and can be also 
called Okubo relation which has been shown by 
Okubo long ago $^{[17]}$ and by Morpurgo recently$^{[18]}$. Here such 
conclusion is again obtained in the skyrmion 
formalism. Apparently, eqs.(~\ref{decupeq1})-(~\ref{decupeq4}) indicate 
that the equal spacing rule
for decuplet (i.e.,GOR in decuplet section) no longer holds. Then a
correction to the standard GOR will exist. That is, the following 
corrections to $\delta m_{10}^{(1)}$ and 
$\delta m_{10}^{(2)}$ (see eq.(1)) can be obtained from 
eq.(~\ref{decuporder2}) by comparing with eq.(1)
\begin{eqnarray}
\label{decup1}
(\delta m_{10}^{(1)})_{second}&=& 2M_{\Sigma^{*}}-M_{\Xi^{*}}-M_{\Delta}.\\
\label{decup5}
(\delta m_{10}^{(2)})_{second}&=& 2(2M_{\Sigma^{*}}-M_{\Xi^{*}}-M_{\Delta}).
\end{eqnarray}
Note that $(\delta m_{10}^{(2)})_{second}=2 (\delta m_{10}^{(1)})_{second}$. 
In practice, this is a general fact at the second order. Meanwhile,
the mass-variables in eq.(~\ref{decup1})(~\ref{decup5}) could be changed, then
the other three expressions for $(\delta m_{10}^{(1)})_{second}$,
$(\delta m_{10}^{(2)})_{second}$ read
\begin{eqnarray}
\label{decup2}
(\delta m_{10}^{(1)})_{second} &=&\frac{1}{3}(3M_{\Xi^{*}}-2M_{\Omega}
-M_{\Delta}),\\
\label{decup3}
   &=&\frac{1}{3}(3M_{\Sigma^{*}}-2M_{\Delta}-M_{\Omega}),  \\
\label{decup4}
   &=&2M_{\Xi^{*}}-M_{\Sigma_{*}}-M_{\Omega}.\\
\label{decup6}
(\delta m_{10}^{(2)})_{second} &=& 2 (\delta m_{10}^{(1)})_{second}
\end{eqnarray}
in eq.(~\ref{decup6}), we could see $(\delta m_{10}^{(2)})_{second}$ 
takes the double value of 
$(\delta m_{10}^{(1)})_{second}$ for each case.

The above four $(\delta m_{10}^{(1)})_{second}$'s are equivalent in principle. 
However, since eq.(~\ref{decuporder2}) is just an approximate relation, 
the numerical
values of the four $(\delta m_{10}^{(1)})_{second}$ are not exactly equal, 
a small error will be given. Therefore we could have
\begin{equation}
\label{decup2nd1}
(\delta m_{10}^{(1)})_{second}= 6.8\pm 2.6 MeV
\end{equation}   
\begin{equation}
\label{decup2nd2}
(\delta m_{10}^{(2)})_{second}=2(\delta m_{10}^{(1)})_{second}=13.6\pm 5.2 MeV,
\end{equation}
which agree well with the experimental data in eqs.(3) and (4).

As for the third order corrections, a constrained mass relation 
similar to eq.(~\ref{decuporder2}) can be obtained, which 
is rather complicated because eqs.(~\ref{decupeq1})-(~\ref{decupeq4}) 
are non-linear.
Therefore, the numerical analysis and comparison turned out to be necessary. 
For the sake of convenience and straightforwardness, the numerical 
values of the baryon masses in decuplet obtained from the
constrained mass relation up to the third order are calculated,
which are listed in table {\bf 2} in comparison with those of 
the first and second order. We only list the predictions for 
$M_{\Omega}$, while $M_{\Delta}, M_{\Sigma^{*}}$ and $M_{\Xi^{*}}$ 
are input. It is easy to see that the theoretical predictions of 
baryon masses agree well with the experiment quantitatively.

Now we study the corrections to $\delta m_{10}^{(1)}$ and 
$\delta m_{10}^{(2)}$ in the third order. Then we could 
have the following equations 
\begin{equation}
M_{\Sigma^{*}}-M_{\Delta}=M_{\Xi^{*}}-M_{\Sigma^{*}}+
(\delta m_{10}^{(1)})_{second}+(\delta m_{10}^{(1)})_{third}
\end{equation}
\begin{equation}
M_{\Sigma^{*}}-M_{\Delta}=M_{\Omega}-M_{\Xi^{*}}+
(\delta m_{10}^{(2)})_{second}+(\delta m_{10}^{(2)})_{third}
\end{equation}
where $(\delta m_{10}^{(1)})_{second}$ and $(\delta m_{10}^{(2)})_{second}$
have been obtained in eqs.(~\ref{decup1})-(~\ref{decup6}) and 
eqs.(~\ref{decup2nd1})-(~\ref{decup2nd2}). Through numerical 
calculations,
$(\delta m_{10}^{(1)})_{third}$ and $(\delta m_{10}^{(2)})_{third}$ 
are found to be 
\begin{equation}
(\delta m_{10}^{(1)})_{third} = -0.5\pm 0.5 MeV.
\end{equation}
\begin{equation}
(\delta m_{10}^{(2)})_{third} = -0.3\pm 0.7 MeV.
\end{equation}
then we get the sums of the corrections of the second and the 
third order as follows,
\begin{equation}
(\delta m_{10}^{(1)})_{sum} = 6.3\pm 3.1 MeV.
\end{equation}
\begin{equation}
(\delta m_{10}^{(2)})_{sum} = 13.3\pm 5.9 MeV.
\end{equation}

It is easy to see that, 
the corrections of the third order are far less than those of 
the second order. But the corrections of the third order 
improve the results of $\delta m_{10}^{(1)}$ and $\delta m_{10}^{(2)}$.
Therefore the results up to the third order are still encouraging.
\vskip0.2in
\centerline{{\bf Table 2. Comparison between higher corrections and the
experimental data for decuplet}}
\begin{center}
\begin{tabular}{|c|cccc|} \hline 
baryons & $M_{\Delta (input)}$ & $M_{\Sigma^{*} (input)}$ & $M_{\Xi^{*}
(input)}$ & $M_{\Omega}$\\ \hline
&&&&\\
 1-st Order & 1232.0 & 1384.6 & 1533.4 & 1684.1 \\
&&&&\\
 2-nd Order & 1232.0 & 1384.6 & 1533.4 & 1678.2 \\ 
&&&&\\
 3-rd Order & 1232.0 & 1384.6 & 1533.4 & 1678.1 \\
&&&&\\
Exper. data & 1232.0 & 1384.6 & 1533.4 & 1672.4 \\
&&&&\\ \hline
\end{tabular}
\end{center}
\vskip0.3in
\subsection{The baryon-octet sector}

 Note that
\begin{equation}
8 \otimes 8=27\oplus 10\oplus 10^{*}\oplus 8_{F}\oplus 8_{D}\oplus 1 
\end{equation}
we have (see eq.$(~\ref{Ek2})\&(~\ref{Ek3})$)
\begin{equation}
\{|m \rangle,|n \rangle\} \in \{ 27, 10, 10^{*}, 1 \}.
\end{equation}
Employing the analysis similar to decuplet, we could obtain the non-zero
matrix elements for octet (see Table {\bf 3}.)
\newpage
\centerline{{\bf Table 3. Values of non-zero matrix elements of 
$H'_{m,n}$ for Octet}}

\begin{center}
\begin{tabular}{|c|cccccc|} \hline 
baryons & $H'_{8,8}/m$ & $H'_{10^{*},10^{*}}/m$ & $H'_{27,27}/m$ 
& $H'_{8,10^{*}}/m$ & $H'_{8,27}/m$ & $H'_{10^{*},27}/m$ \\ \hline
&&&&&&\\
N & $\frac{7}{10}$ & $\frac{7}{8}$ & $\frac{423}{560}$ & 
$\frac{\sqrt{5}}{10}$ & $\frac{\sqrt{6}}{10}$ & 
$\frac{\sqrt{30}}{80}$\\
&&&&&&\\
$\Sigma$ & $\frac{11}{10}$ & 1 & $\frac{267}{280}$ & 
$\frac{\sqrt{5}}{10}$ & $\frac{1}{5}$ & 
$\frac{\sqrt{5}}{10}$\\
&&&&&&\\
$\Lambda$ & $\frac{9}{10}$ & 0 & $\frac{57}{70}$ & 
0 & $\frac{3}{10}$ & 0\\
&&&&&&\\
$\Xi$ & $\frac{6}{5}$ & 0 & $\frac{33}{35}$ & 0 & $\frac{\sqrt{6}}{10}$ & 0\\ 
&&&&&&\\ 
\hline
\end{tabular}
\end{center}

Noticing $C_{2}(10^{*})=6$ and making use of eq.(~\ref{Ek0-En0}), we get
\begin{eqnarray}
E_{8}^{(0)}-E_{10^{*}}^{(0)} &=&-\frac{3}{2b^{2}} \\
E_{8}^{(0)}-E_{27}^{(0)} &=&-\frac{5}{b^{2}}
\end{eqnarray}
Similarly we could obtain the complete results for baryons in octet
up to the third order{\footnote{In ref.[6], the numerical 
value of the third order for $M_{\Sigma}$ is incorrect.}}
\begin{eqnarray}
\label{octeteq1}
M_{N} &=& M_{8} - \frac{3}{10} m -\frac{43}{750} m^{2}b^{2}
	+\frac{3812}{1575000} m^{3} b^{4}\\
\label{octeteq2}
M_{\Lambda} &=& M_{8} -\frac{1}{10} m - \frac{9}{250} m^{2}b^{2} 
	-\frac{54}{43750} m^{3} b^{4}\\
\label{octeteq3}
M_{\Sigma} &=& M_{8} + \frac{1}{10} m -\frac{37}{750} m^{2}b^{2}
	-\frac{1672}{196875} m^{3} b^{4}\\
\label{octeteq4}
M_{\Xi} &=& M_{8} + \frac{1}{5} m -\frac{3}{125} m^{2}b^{2} 
	-\frac{54}{21875} m^{3} b^{4}
\end{eqnarray}
where $M_{8}=\langle H_{0} \rangle _{\lambda =8}+m$. Similar to the
decuplet, the mass relation for the baryons in octet could be obtained.

At the first order, the standard GOR for octet can be easily got
\begin{equation}
\label{octetorder1}
2(M_{N}+M_{\Xi})=3M_{\Lambda}+M_{\Sigma},\\
\end{equation}

At the second order, a modified mass relation for baryons in octet
can be obtained
\begin{equation}
\label{octetorder2}
15 M_{\Sigma}+35 M_{\Lambda}=24 M_{N}+26 M_{\Xi}
\end{equation}
which is the IGOR for octet. Repeating the procedure for decuplet and
comparing with eq.(2), four forms of corrections to
$\delta m_{8}$ for octet can be obtained from eq.(~\ref{octetorder2})
\begin{eqnarray}
\label{oct1}
(\delta m_{8})_{second} &=& \frac{2}{13}(M_{N}+M_{\Sigma}-2M_{\Lambda}), \\
\label{oct2}
	    &=& \frac{1}{12}(3M_{\Sigma}-2M_{\Xi}-M_{\Lambda}), \\
\label{oct3}
	    &=& \frac{2}{35}(5M_{\Sigma}-M_{N}-4M_{\Xi}), \\
\label{oct4}
	    &=& \frac{1}{15}(6M_{N}+4M_{\Xi}-10M_{\Lambda}).  
\end{eqnarray}
The above four $(\delta m_{8})_{second}$ are also equivalent in principle. 
However, since eq.(~\ref{octetorder2}) is still not an exact identity of 
masses, the numerical results of $(\delta m_{8})_{second}$ given by 
eqs.(~\ref{oct1})-(~\ref{oct4}) 
will give the theoretical prediction with an error as follows
\begin{equation}
\label{oct2nd}
(\delta m_{8})_{second}=-15.1 \pm 1.2 MeV
\end{equation}
Obviously, the second order corrections to $\delta m_{8}$ agree well
with experiment quantitatively. 

As for the third order corrections, we will go on to carry out the 
numerical analysis and the comparison. The procedures are also similar 
to the decuplet and the results are listed in Table {\bf 4} in comparison 
with those of the first and second order. Similar to the decuplet, we
only list prediction for the mass of $M_{\Xi}$
while others are input. It could be easily 
seen from Table {\bf 4} that the results at the second order 
agree with experiment fairly well, however it is not the case
for the third order: the predictions go far away from the experiment 
in comparison with the second order. This can be shown more
directly by studying the corrections up to the third order.

Similar to the procedure of decuplet, the sum of the corrections 
to $\delta m_{8}$ in eq.(~\ref{GOR8}) up to the third order can be
writen as
\begin{equation}
2(M_{N}+M_{\Xi})=3M_{\Lambda}+M_{\Sigma}+(\delta m_{8})_{second}
+(\delta m_{8})_{third}
\end{equation}
where $(\delta m_{8})_{second}$ have been obtained in 
eqs.(~\ref{oct1})-(~\ref{oct4}) and eq.(~\ref{oct2nd}). 
Through numerical calculations, $(\delta m_{8})_{third}$ 
is found to be 
\begin{equation}
(\delta m_{8})_{third} = 60.2\pm 19.1 MeV.
\end{equation}
then we get the sum of the corrections of the second and the 
third order as follows,
\begin{equation}
(\delta m_{8})_{sum} = 45.1\pm 20.3 MeV.
\end{equation}
Obvoiusly, the corrections are much far away from the actual deviation
as is shown in eq.(5). In Fig.1, we visually display the predictions
for $M_{\Omega}$ in decuplet and $M_{\Xi}$ in octet
between the different order corrections.
\vskip0.2in
\centerline{{\bf Table 4. Comparison between higher corrections and the
experimental data for octet}}
\begin{center}
\begin{tabular}{|c|cccc|} \hline 
baryons & $M_{N (input)}$ & $M_{\Lambda (input)}$ & $M_{\Sigma (input)}$ 
& $M_{\Xi}$\\ \hline
&&&&\\
 1-st Order & 938.9 & 1115.6 & 1193.0 & 1331.1 \\
&&&&\\
 2-nd Order & 938.9 & 1115.6 & 1193.0 & 1323.4 \\ 
&&&&\\
 3-rd Order & 938.9 & 1115.6 & 1193.0 & 1366.2 \\
&&&&\\
Exper. data & 938.9 & 1115.6 & 1193.0 & 1318.0 \\
&&&&\\ \hline
\end{tabular}
\end{center}

\newpage
\setlength{\unitlength}{0.1cm}
\begin{picture}(140,170)
\put(15,15){\thicklines \line(1,0){30}}
\put(15,15.15){\thicklines \line(1,0){30}}
\put(15,35){\line(1,0){30}}
\put(15,36){\line(1,0){30}}
\put(15,57){\line(1,0){30}}
\put(20,15){\vector(0,1){42}}
\put(30,15){\vector(0,1){21}}
\put(40,15){\vector(0,1){20}}
\put(30,7){\makebox(0,0)[c]{$M_{\Omega}$(decuplet)}}
\put(5,15){\makebox(0,0)[c]{Expt.}}
\put(3,33.5){\makebox(0,0)[c]{3-rd Order}}
\put(3,37){\makebox(0,0)[c]{2-nd Order}}
\put(3,60){\makebox(0,0)[c]{1-st Order}}
\put(16,40){\makebox(0,0)[c]{11.7}}
\put(27,25){\makebox(0,0)[c]{5.8}}
\put(37,25){\makebox(0,0)[c]{5.7}}
\put(100,15){\thicklines \line(1,0){30}}
\put(100,15.15){\thicklines \line(1,0){30}}
\put(100,34){\line(1,0){30}}
\put(100,55){\line(1,0){30}}
\put(100,145){\line(1,0){30}}
\put(105,15){\vector(0,1){40}}
\put(115,15){\vector(0,1){19}}
\put(125,15){\vector(0,1){130}}
\put(115,7){\makebox(0,0)[c]{$M_{\Xi}$(octet)}}
\put(90,15){\makebox(0,0)[c]{Expt.}}
\put(88,55){\makebox(0,0)[c]{1-st Order}}
\put(88,34){\makebox(0,0)[c]{2-nd Order}}
\put(88,145){\makebox(0,0)[c]{3-rd Order}}
\put(100,40){\makebox(0,0)[c]{11.1}}
\put(112,25){\makebox(0,0)[c]{5.4}}
\put(120,70){\makebox(0,0)[c]{48.2}}
\end{picture}
\begin{description}
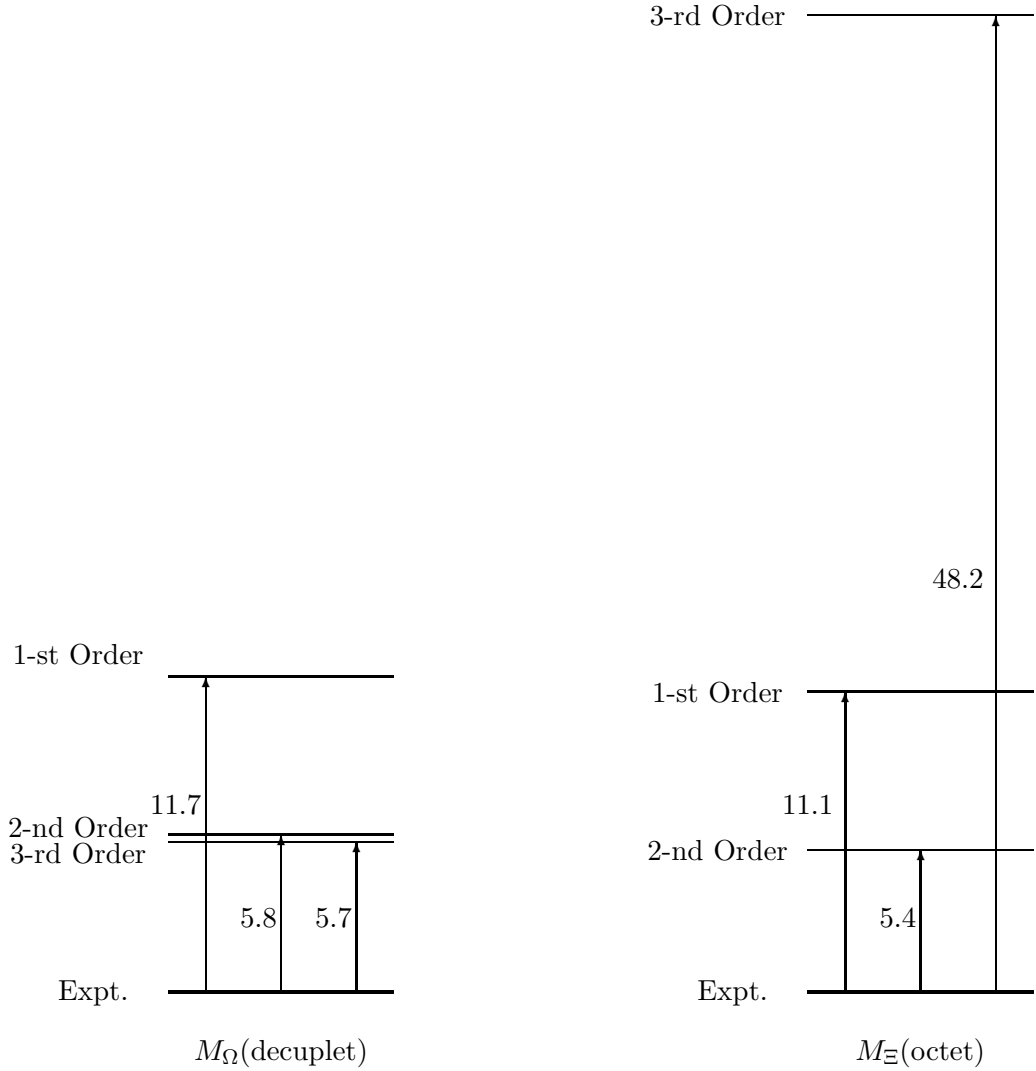
\item[Fig.1] The predictions of the mass of $M_{\Omega}$ in decuplet 
and $M_{\Xi}$ in octet in different orders. $M_{\Delta},
M_{\Sigma}^{*},M_{\Xi}^{*}$ and $M_{N},M_{\Lambda},M_{\Sigma}$ 
are input and the unit is $MeV$.
\end{description}

\section{Comments and Conclusions}

 In the above text, the higher order corrections to GOR's are 
investigated by means of perturbation in the 
framework of standard SU(3) Skyrme model.
The results are found to be quite compatible with experiment 
up to the second order. However, things gets complicated at 
the third order. For decuplet, the results become further 
encouraging though the improvements are not very remarkable. 
Yet for octet, the results become quite disagreable. 
This means that, at the perturbation level, the calculations of
corrections to GOR in the standard SU(3) Skyrme model can hold 
to the third order for decuplet and only to the second order 
for octet-baryons. 

As we know, the key point of the corrections to GOR's is the 
calculations of the non-zero matrix elements (see eqs.(~\ref{Ek0})
-(~\ref{Ek3}) and eq.(~\ref{matrixelement})), which are 
closely related to the flavor breaking term. Then it could 
be seen that there exist defects in present standard flavor 
breaking term.

The problem has been discussed by many authors$^{[8],[16],[19,20]}$. 
Among these discussions, in our opinion, one reasonable scheme 
was put forward by one of us in ref.[16,21]. In that model,
the chiral SU(3) flavor group is embeded into a larger 
SU($N_{f}$) group, and the (ud)-(s) flavor symmetry breaking 
term still keeps the standard form like eq.(~\ref{H'}). 
Basically, in the chiral Lagrangian theory from QCD, neglecting
the ${\theta}$-dependence, the standard flavor symmetry breaking 
term of eq.(~\ref{H'}) comes directly from the following quark 
mass term $^{[22]}$
\begin{equation}
{\cal L}_{m} = \frac{1}{8} \frac{m_{\pi}^{2} F_{\pi}^{2}}{m_{u} + m_{d}}
Tr[M_{q}(U+U^{\dag}-2)]
\end{equation}
where $M_q$=diag($m_u,m_d,m_s$).${\cal L}_m$ is also directly
related to PCAC (partial conservation of axial vector current).
In the sub-SU(3) Skyrme model, only such symmetry breaking 
mechanism is adopted and no non-standard terms with adjustable
parameters are introduced. Therefore its success is expected
and reasonable.
On the other hand, it was found that the mass spectra for decuplet 
and octet are quite compatible with the experimental data at the 
first order as $N = 5 \sim 6 $. Apparently, this is interesting: 
in the real world, there are only six known quark flavors.
Then we may speculate this idea may be possible to completely 
account for the deviations between GOR and the experimental data.
Therefore it is still of great need to carry out the higher 
order perturbative calculations in the framework of ref.[16].
However such an investigation is out of the scope of this 
paper due to the lack of enough C-G coefficients for $SU(N_f)$ 
with $N_f>3$ at present, and will be left to the further studies.

Finally we want to make some comments on the rotating excited states.
As a puzzle, it was attended long ago whether there exist some ``exotic''
SU(3) baryon-multiplets besides the ordinary octet 
and decuplet in the flavor SU(3) theory. In this paper, 
we could see from the procedure of the calculations
that the rotating excited states have played 
important role in the improvement of baryon mass relations, 
since the higher order perturbation could be attributed to 
the calculations of the $H^{\prime}$-matrix elements between 
various eigenstates of $H_{0}$
(see eqs.(~\ref{Eseries})-(~\ref{Ek3})).
So the above results could be regarded as signals of the existence 
of such rotating excited SU(3)-multiplets as 10$^*$-let (with spin 
$\frac{1}{2}$), 27-let (with spin $\frac{1}{2}$ or $\frac{3}{2}$) 
and 35-let (with spin $\frac{3}{2}$). Meanwhile, the wave-functions 
of these SU(3)-states satisfy the constrained condition coming from 
the Wess-Zumino term in the QCD effective Lagrangian, i.e., the 
spin-hypercharge $Y_{R}=1 ^{[9]}$. So they should be physical states 
in QCD. 
However, up to now, no any experimental evidences have indicated 
the existence of these states. A possible explanation seems to be the 
unusually large width of the resonances which makes the identification 
very difficult in experiment.
However, it is just because that these rotating excited states contribute 
to the deviations of GOR from the experiment by the method of perturbative 
calculations, we could ``see'' them indirectly through the corrections
to $\delta m_{10}^{(1)}, \delta m_{10}^{(2)}, \delta m_{8}$. 
So the above positive conclusion is of course interesting.
\newpage
\centerline{\Large Appendix A: Tables for Isoscalar factors of 
$8\otimes35\rightarrow 35$}
\vskip0.2in

The tables in this Appendix are about the SU(3) iso-scalar factors:
\begin{eqnarray*}
\left (\begin{array}{cccc}
\lambda_{1} & \lambda_{2} & | & \lambda \\
I_{1}Y_{1} & I_{2} Y_{2} & | & I Y 
\end{array} \right) \hspace{3.7in} {(A.1)}
\end{eqnarray*}
First of all, we want to present the phase conventions adopted 
in this paper, because some of them are different from the 
corresponding ones in the ref.[13].

The phase conventions are adopted as follows:

1. The relative phase conventions within a definite iso-multiplet are 
determined by the Condon and Shortley phase convention. i.e.,
\begin{eqnarray*}
I_{+} \phi(I,I_{z},Y) &=& [(I-I_{z})(I+I_{z}+1)]^{\frac{1}{2}} 
\phi(I,I_{z}+1,Y), \hspace{1.5in}{(A.2)}\\
I_{-} \phi(I,I_{z},Y) &=& [(I+I_{z})(I-I_{z}+1)]^{\frac{1}{2}} 
\phi(I,I_{z}-1,Y), \hspace{1.5in}{(A.3)}
\end{eqnarray*}

2. The relative phase between the different iso-multiplets
are defined in eq.(~\ref{basis}) (see Sec.3). This is different from 
that of de Swart's$^{[13]}$. 

3. In the CG series $m_{1} \otimes m_{2} \rightarrow \sum_{i} n_{i} $, 
the relative phase between $n_{i}$ and $m_{1},m_{2}$ are determined in 
this way: 

First, we set all the phase factors always  real.
Second, we determine the sign of the phase by considering the highest
eigenstates $\phi_{\mu}^{\lambda}$ of the IR's $n_{i}$

\begin{eqnarray*}
\phi_{\mu}^{\lambda}=\sum_{\scriptsize 
I_{1},Y_{1},I_{1z},I_{2},Y_{2},I_{2z} }
\left (\begin{array}{cccc}
 \lambda_{1} & \lambda_{2} & | & \lambda \\
I_{1} Y_{1} & I_{2} Y_{2} & | & I Y 
\end{array} \right) C_{CG}[SU(2)]~ \phi_{\mu_{1}}^{\lambda_{1}} 
\phi_{\mu_{2}}^{\lambda_{2}}\hspace{1.3in}{(A.4)}
\end{eqnarray*}

Among the different SU(3) iso-scalar factors, we choose the one with
the largest possible $I_{1}$ to be positive. If this is not sufficient,
we take from the iso-scalar factors with the largest possible $I_{1}$
the one with the largest possible $I_{2}$ positive.

The conventions here are sufficient to decide all phases factors in SU(3). 
Conventions 2 and 3 are different from those of ref.[13]. 
Convention 2 is easy to understand, since we do not use $V$-spin (in ref.[13]'s
notation, it is denoted with $K$.), but Gel'fand basis technique.
Covention 3 is a supplement to ref.[13], since the corresponding one 
is not sufficient in ref.[13]. 

All the CGC's (or iso-scalar factors) that are related to the calculations
in this paper have been re-calculated according to the above 
conventions by means of Gel'fand basis. The results show that the CGC's
by employing Gel'fand basis and the above conventions are completely 
consistent with the CGC's in ref.[13]. So we could use alternatively 
both CGC's in ref.[13] and those obtained by the technique in the 
present paper without any contradiction.

As for  SU(N) with $N \geq 4$, new phase conventions are needed. In principle,
they can be introduced according to the above method.

\vskip0.4in
\centerline{Table A.1: Y=2, I=2}

\begin{center}
\tabcolsep0.2in
\begin{tabular}{|ll|ll|c|c|}\hline
$Y_{1}$ & $I_{1}$ & $Y_{2}$ & $I_{2}$ & $35_{1}$ & $35_{2}$ \\ \hline
1 & $\frac{1}{2}$ & 1 & $\frac{5}{2}$ & 0 & $-\sqrt{\frac{27}{40}}$ \\
1 & $\frac{1}{2}$ & 1 & $\frac{3}{2}$ & $-\sqrt{\frac{5}{27}}$ &$
\sqrt{\frac{2}{135}}$ \\
0 & 1 & 2 & 2 & $\sqrt{\frac{2}{3}}$ &$\sqrt{\frac{1}{12}}$\\
0 & 0 & 2 & 2 & $\sqrt{\frac{4}{27}}$ &$-\sqrt{\frac{49}{216}}$\\ \hline
\end{tabular}
\end{center}
\vskip0.3in
\centerline{Table A.2: Y=1, I=$\frac{5}{2}$}

\begin{center}
\tabcolsep0.2in
\begin{tabular}{|ll|ll|c|c|}\hline
$Y_{1}$ & $I_{1}$ & $Y_{2}$ & $I_{2}$ & $35_{1}$ & $35_{2}$ \\ \hline
1 & $\frac{1}{2}$ & 0 & 2 & $-\sqrt{\frac{2}{9}}$ & $-\frac{1}{6}$ \\
0 & 1 & 1 & $\frac{5}{2}$ & $\sqrt{\frac{140}{225}}$ & 
$-\sqrt{\frac{175}{1440}}$\\
0 & 1 & 1 & $\frac{3}{2}$ & $\sqrt{\frac{1}{135}}$ &$\sqrt{\frac{30}{324}}$\\ 
0 & 0 & 1 & $\frac{5}{2}$ & $\sqrt{\frac{4}{27}}$ &
$\sqrt{\frac{169}{864}}$\\ 
$-1$ & $\frac{1}{2}$ & 2 & 2 & 0 & $-\frac{3}{4}$ \\ \hline
\end{tabular}
\end{center}
\vskip0.3in
\centerline{Table A.3: Y=1, I=$\frac{3}{2}$}

\begin{center}
\tabcolsep0.2in
\begin{tabular}{|ll|ll|c|c|}\hline
$Y_{1}$ & $I_{1}$ & $Y_{2}$ & $I_{2}$ & $35_{1}$ & $35_{2}$ \\ \hline
1 & $\frac{1}{2}$ & 0 & 2 & $\frac{\sqrt{3}}{36}$ &$-\sqrt{\frac{121}{216}}$ \\
1 & $\frac{1}{2}$ & 0 & 1 & $-\sqrt{\frac{125}{432}}$ &
$\sqrt{\frac{5}{216}}$ \\
0 & 1 & 1 & $\frac{5}{2}$ & $-\sqrt{\frac{1}{90}}$ &$-\sqrt{\frac{5}{36}}$\\
0 & 1 & 1 & $\frac{3}{2}$ & $\sqrt{\frac{961}{2160}}$ 
&$-\sqrt{\frac{30}{324}}$\\
0 & 0 & 1 & $\frac{3}{2}$ & $\sqrt{\frac{1}{48}}$ &$-\sqrt{\frac{1}{6}}$ \\
$-1$ & $\frac{1}{2}$ & 2 & 2 & $\sqrt{\frac{25}{108}}$ &$-\sqrt{\frac{1}{54}}$\\
\hline
\end{tabular}
\end{center}

\newpage
\vskip0.5in
\centerline{Table A.4: Y=0, I=2}

\begin{center}
\tabcolsep0.2in
\begin{tabular}{|ll|ll|c|c|}\hline
$Y_{1}$ & $I_{1}$ & $Y_{2}$ & $I_{2}$ & $35_{1}$ & $35_{2}$ \\ \hline
1 & $\frac{1}{2}$ & $-1$ & $\frac{3}{2}$ & $-\sqrt{\frac{1}{3}}$ &
 $-\sqrt{\frac{1}{24}}$ \\
0 & 1 & 0 & 2 & $\sqrt{\frac{9}{24}}$ & $-\sqrt{\frac{3}{16}}$ \\
0 & 1 & 0 & 1 & $\sqrt{\frac{1}{72}}$ & $\frac{5}{12}$\\
0 & 0 & 0 & 2 & $\sqrt{\frac{1}{108}}$ &$\sqrt{\frac{25}{216}}$\\ 
$-1$ & $\frac{1}{2}$ & 1 & $\frac{5}{2}$ & $\sqrt{\frac{4}{15}}$ &
 $\sqrt{\frac{1}{30}}$ \\
$-1$ & $\frac{1}{2}$ & 1 & $\frac{3}{2}$ & $\frac{\sqrt{15}}{90}$ &
 $-\frac{11}{90}\sqrt{30}$ \\ \hline
\end{tabular}
\end{center}
\vskip0.5in
\centerline{Table A.5: Y=0, I=1}

\begin{center}
\tabcolsep0.2in
\begin{tabular}{|ll|ll|c|c|}\hline
$Y_{1}$ & $I_{1}$ & $Y_{2}$ & $I_{2}$ & $35_{1}$ & $35_{2}$ \\ \hline
1 & $\frac{1}{2}$ & $-1$ & $\frac{3}{2}$ & $\frac{1}{9}$ & 
$-\frac{17}{36}\sqrt{2}$ \\
1 & $\frac{1}{2}$ & $-1$ & $\frac{1}{2}$ & $-\frac{5}{9}$ &
$\sqrt{\frac{2}{81}}$ \\
0 & 1 & 0 & 2 & $-\frac{1}{4}\sqrt{\frac{10}{27}}$ & 
$-\frac{5}{36}\sqrt{15}$\\
0 & 1 & 0 & 1 & $\frac{13}{36}\sqrt{2}$ & $\frac{11}{36}$ \\ 
0 & 0 & 0 & 1 & $-\frac{\sqrt{3}}{18}$ & $-\frac{5}{36}\sqrt{6}$\\
$-1$ & $\frac{1}{2}$ & 1 & $\frac{3}{2}$ & $\frac{5}{18}\sqrt{5}$ & 
$-\frac{\sqrt{10}}{18}$\\ \hline
\end{tabular}
\end{center}
\vskip0.5in
\centerline{Table A.6: Y=-1, I=$\frac{3}{2}$}

\begin{center}
\tabcolsep0.2in
\begin{tabular}{|ll|ll|c|c|}\hline
$Y_{1}$ & $I_{1}$ & $Y_{2}$ & $I_{2}$ & $35_{1}$ & $35_{2}$ \\ \hline
1 & $\frac{1}{2}$ & $-2$ & 1 & $-\sqrt{\frac{1}{3}}$ & $-\sqrt{\frac{1}{24}}$ \\
0 & 1 & $-1$ & $\frac{3}{2}$ & $\sqrt{\frac{5}{27}}$ & $-\sqrt{\frac{245}{864}}$\\
0 & 1 & $-1$ & $\frac{1}{2}$ & $\sqrt{\frac{1}{54}}$ & $\sqrt{\frac{75}{324}}$\\
0 & 0 & $-1$ & $\frac{3}{2}$ & $-\sqrt{\frac{1}{27}}$& $\sqrt{\frac{49}{864}}$\\
$-1$ & $\frac{1}{2}$ & 0 & 2 & $\sqrt{\frac{15}{36}}$ & $\sqrt{\frac{30}{576}}$\\
$-1$ & $\frac{1}{2}$ & 0 & 1 & $\sqrt{\frac{1}{108}}$ & 
$-\frac{17}{72}\sqrt{6}$\\
\hline
\end{tabular}
\end{center}
\newpage
\vskip0.16in
\centerline{Table A.7: Y=-1, I=$\frac{1}{2}$}

\begin{center}
\tabcolsep0.18in
\begin{tabular}{|ll|ll|c|c|}\hline
$Y_{1}$ & $I_{1}$ & $Y_{2}$ & $I_{2}$ & $35_{1}$ & $35_{2}$ \\ \hline
1 & $\frac{1}{2}$ & $-2$ & 1 & $\sqrt{\frac{1}{24}}$ & $-\sqrt{\frac{1}{3}}$ \\
1 & $\frac{1}{2}$ & $-2$ & 0 & $-\sqrt{\frac{25}{108}}$ & $\sqrt{\frac{1}{54}}$\\
0 & 1 & $-1$ & $\frac{3}{2}$ & $-\sqrt{\frac{1}{27}}$ & $\sqrt{\frac{25}{54}}$\\
0 & 1 & $-1$ & $\frac{1}{2}$ & $\sqrt{\frac{49}{432}}$ &$\sqrt{\frac{2}{27}}$\\
0 & 0 & $-1$ & $\frac{1}{2}$ & $-\sqrt{\frac{49}{432}}$ &$-\sqrt{\frac{2}{27}}$\\
$-1$ & $\frac{1}{2}$ & 0 & 1 & $\sqrt{\frac{25}{54}}$ & 
$-\sqrt{\frac{1}{27}}$\\ \hline
\end{tabular}
\end{center}
\vskip0.16in
\centerline{Table A.8: Y=-2, I=1}

\begin{center}
\tabcolsep0.2in
\begin{tabular}{|ll|ll|c|c|}\hline
$Y_{1}$ & $I_{1}$ & $Y_{2}$ & $I_{2}$ & $35_{1}$ & $35_{2}$ \\ \hline
1 & $\frac{1}{2}$ & $-3$ & $\frac{1}{2}$ & $-\sqrt{\frac{2}{9}}$ &
 $-\frac{1}{6}$ \\
0 & 1 & $-2$ & 1 & $\sqrt{\frac{1}{18}}$ &$-\frac{2}{3}$ \\
0 & 1 & $-2$ & 0 & $\sqrt{\frac{1}{54}}$ &$\sqrt{\frac{25}{108}}$\\
0 & 0 & $-2$ & 1 & $-\sqrt{\frac{25}{108}}$ &$\sqrt{\frac{1}{54}}$\\
$-1$ & $\frac{1}{2}$ & $-1$ & $\frac{3}{2}$ & $\frac{2}{3}$ &
$\sqrt{\frac{1}{18}}$ \\
$-1$ & $\frac{1}{2}$ & $-1$ & $\frac{1}{2}$ & $\frac{1}{6}$ &
$-\sqrt{\frac{2}{9}}$ \\ \hline
\end{tabular}
\end{center}
\vskip0.16in
\centerline{Table A.9: Y=-2, I=0}

\begin{center}
\tabcolsep0.2in
\begin{tabular}{|ll|ll|c|c|}\hline
$Y_{1}$ & $I_{1}$ & $Y_{2}$ & $I_{2}$ & $35_{1}$ & $35_{2}$ \\ \hline
1 & $\frac{1}{2}$ & $-3$ & $\frac{1}{2}$ & $\sqrt{\frac{4}{27}}$ & 
$-\sqrt{\frac{49}{216}}$ \\
0 & 1 & $-2$ & 1 & $-\sqrt{\frac{1}{18}}$ & $-\frac{5}{6}$ \\
0 & 0 & $-2$ & 0 & $-\sqrt{\frac{1}{3}}$ & $-\sqrt{\frac{1}{24}}$\\
$-1$ & $\frac{1}{2}$ & $-1$ & $\frac{1}{2}$ & $\sqrt{\frac{50}{108}}$ &
$-\sqrt{\frac{1}{27}}$\\ \hline
\end{tabular}
\end{center}
\vskip0.16in
\centerline{Table A.10: Y=-3, I=$\frac{1}{2}$}

\begin{center}
\tabcolsep0.2in
\begin{tabular}{|ll|ll|c|c|}\hline
$Y_{1}$ & $I_{1}$ & $Y_{2}$ & $I_{2}$ & $35_{1}$ & $35_{2}$ \\ \hline
0 & 1 & $-3$ & $\frac{1}{2}$ & 0 &$-\sqrt{\frac{81}{96}}$\\
0 & 0 & $-3$ & $\frac{1}{2}$ & $-\sqrt{\frac{16}{27}}$ &
$\sqrt{\frac{1}{864}}$\\ 
$-1$ & $\frac{1}{2}$ & $-2$ & 1 & $\sqrt{\frac{1}{3}}$ & 
$\sqrt{\frac{1}{24}}$\\
$-1$ & $\frac{1}{2}$ & $-2$ & 0 & $\sqrt{\frac{2}{27}}$ &
$-\sqrt{\frac{49}{432}}$\\
\hline
\end{tabular}
\end{center}

\end{document}